\documentclass[twocolumn,showpacs,showkeys,preprintnumbers,amsmath,amssymb,superscriptaddress,nofootinbib,prl]{revtex4}
\usepackage{dcolumn}
\usepackage{bm}
\usepackage{graphics}
\usepackage{graphicx}

\begin{document}

\title{Comment on the possibility of Inverse Crystallization within \\ van Hemmen's Classical Spin-Glass model}

\author{Eduardo Cuervo-Reyes}
\affiliation{Laboratory of Inorganic Chemistry, Solid State, ETH-Z\"urich 8093 Switzerland}

\date{\today}
%


\maketitle
\begin{figure}
\includegraphics[width=0.9\linewidth]{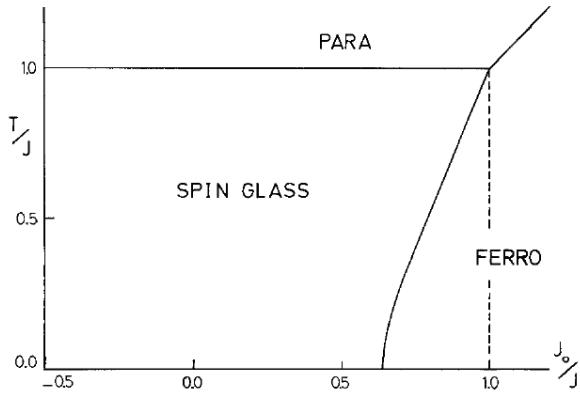}
\caption{Phase diagram of van Hemmen's classical spin  model, at zero external field, with Gaussian-distributed exchange constants; taken from reference\cite{vH2}.}\label{PD}
\end{figure}

In reference\cite{vH}, van Hemmen (vH) introduced a classical spin model with frustration, which is analytically solvable and reproduces many experimental features of a spin glass (SG) quite well\cite{vH}. It consists of $N$ Ising-spins interacting via a long-range random coupling of the form $J^{}_{ij}=J^{}_0/N - (J/N) (\xi_i\eta_j+\xi_j\eta_i)$;  $\xi_i$ and $\eta_j$ are identically distributed random variables with zero mean and finite variance. $J_0$ is a net direct exchange, and  $J$ represents the strength of the strongly-oscillating and long-ranged  Ruderman-Kittel-Kasuya-Yosida (RKKY) interaction. vH phase diagram (Fig.\ref{PD}) is universal\cite{vH2,Celik}; i.e., robust against variations of the distribution of $\xi_i$ and $\eta_j$.   The ferromagnetic state (FE), whenever it exists ($J_0^{}/J>x^{}_c = 2/\pi$), is the low temperature phase; vH considered this is  a limitation of his model.  Apparently, it cannot account for phenomena such as inverse freezing, where an ordered phase appears at higher temperatures than a less-ordered one.  This comment is intended to show that the vH model, in its original classical version, can also account for transitions from a SG phase to a higher-temperature FE state.  

The so far ignored fact is the  temperature dependence of $J$. For $T>0$, the RKKY interaction becomes exponentially damped. The envelope function has the asymptotic form $I(T)=I(0)e^{-\pi T k^{}_{F} r/T^{}_F}$, for $T<T^{}_F$,  irrespective of the dimensionality\cite{Kim}. $T^{}_F$ and $k^{}_{F}$ are the Fermi-temperature and wave-vector, and $r$ is the generic distance between two interacting magnetic centers. As $J$ is representative of the long-range RKKY coupling, this can be roughly estimated  taking the value of the envelope function at $k^{}_{F}r=\pi$. Hence, $J(T)\sim J(0) e^{-\gamma T}$, with $\gamma=\pi^2_{}/T^{}_F$ and $J(0)=J(T=0)$. For a given system, the corresponding $x=J_0^{}/J$ will increase as one goes up in temperature. In other words, a system is represented by fixed $J^{}_0/J(0)$ and $\gamma$; heating/cooling a  sample  does not correspond to a vertical line in Fig.(\ref{PD}). A FE phase may appear upon heating a glass state if $\gamma J(0)\geq x^{-1}_c\ln{(1/x^{}_c)}$; or in terms of $T^{}_F$,
\begin{equation}
J(0)/T^{}_F \geq \pi^{-2}_{}x^{-1}_c\ln{(1/x^{}_c)}.
\end{equation} 

 To check whether this is a reasonable constraint, let use the effective mass approximation. The damping coefficient is  rewritten as $\gamma= 4\pi^2_{} m^*_{}k^{}_B (3n^{}_e/\pi)^{-2/3}_{}/\hbar^2_{}$, in terms of the effective mass $m^*_{}$  and concentration $n^{}_e$ of conduction electrons. $k^{}_B$ and $\hbar$ are the the Boltzmann- and  Plank-constant. Substituting the respective values, we obtain  
\begin{equation}
m^*_{\rm rel}n_e^{-2/3}J(0)\geq 10^{-17}_{}\; {\rm K}\; {\rm m}^2_{}\label{const}
\end{equation}
\noindent where $m^*_{\rm rel}$ is the effective mass relative to the free electron mass. For a good metal ($n^{}_e\sim 10^{28}_{}$ m${}^{-3}_{}$), Eq.(\ref{const}) gives $m^*_{\rm rel}J(0)\geq 50$ K; and for a bad metal or a doped semiconductor ($n^{}_e\sim 10^{22}_{}$ m${}^{-3}_{}$), $m^*_{\rm rel}J(0)\geq 0.005$\nolinebreak K. This is quite reasonable; therefore, vH model may reproduce the inverse crystallization of actual systems. Lastly, it is worth noting that the exponential decay of $J$ can be also used above the tricritical point ($T/J=1;J^{}_0/J=1$), as long as the latter remains below  the Fermi temperature. This will hold if $J(0)/T^{}_F < e^{\pi^2}_{}$, which is satisfied in most reasonable cases. Similarly, taking into account the temperature dependence of the RKKY interaction may change our interpretation of other models.

\end{document}